\documentclass[journal]{IEEEtran}
\IEEEoverridecommandlockouts

\usepackage{cite}
\usepackage{amsmath,amssymb,amsfonts}
\usepackage{mathtools} 
\usepackage{graphicx}
\usepackage{textcomp}
\usepackage{xcolor}
\definecolor{MATLAB_yellow}{rgb}{0, 0.4470, 0.7410}
\definecolor{MATLAB_blue}{rgb}{0.9290, 0.6940, 0.1250}
\definecolor{MATLAB_red}{rgb}{0.8500, 0.3250, 0.0980}
\definecolor{MATLAB_purple}{rgb}{0.4940, 0.1840, 0.5560}
\definecolor{MATLAB_green}{rgb}{0.4660, 0.6740, 0.1880}
\definecolor{MATLAB_azure}{rgb}{0.3010, 0.7450, 0.9330}
\usepackage{tabularx}
\usepackage{booktabs}
\usepackage{float}
\usepackage{algpseudocode}
\usepackage[english]{babel}
\usepackage{graphicx,lipsum}
\usepackage{paralist}
\usepackage{array}
\usepackage{amsmath,stackrel}
\usepackage{subfig}
\usepackage{multicol}
\usepackage{verbatim}
\usepackage{enumerate}              
\usepackage{dsfont}
\usepackage{psfrag}
\usepackage{algorithm}
\usepackage[nolist]{acronym}
\usepackage[utf8]{inputenc}
\usepackage{tikz}
\usepackage{collcell}
\usepackage{pgfplots}
\pgfplotsset{compat=1.18}
\usepackage{hhline}
\usepackage{colortbl}
\usepackage{multido}
\usepackage{multirow}
\usepackage{dblfloatfix}    
\usepackage{etoolbox}
\usepackage{soul}
\usepackage{eucal}
\usepackage[font=footnotesize]{caption}

\begin{acronym}
\acro{AWGN}{additive white Gaussian noise}
\acro{AI}{artificial intelligence}
\acro{AoA}{angle of arrival}
\acro{BF}{beamformer}
\acro{BS}{base station}
\acro{CP}{cyclic prefix}
\acro{CNN}{convolutional neural network}
\acro{CKF}{cubature Kalman filter}
\acro{CPHD}{cardinalized probability hypothesis density}
\acro{DBSCAN}{density-based spatial clustering of applications with noise}
\acro{DoA}{direction of arrival}
\acro{DoD}{direction of departure}
\acro{EIRP}{effective isotropic radiated power}
\acro{ELP}{equivalent low-pass}
\acro{ESPRIT}{Estimation of Signal Parameters via Rotational Invariance Techniques}
\acro{FC}{fusion center}
\acro{FD}{frequency division}
\acro{FFT}{fast Fourier transform}
\acro{FDD}{frequency-division duplexing}
\acro{GM}{Gaussian mixture}
\acro{GMPHD}{Gaussian mixture probability hypothesis density}
\acro{GMCPHD}{Gaussian mixture cardinalized probability hypothesis density}
\acro{GOSPA}{generalized optimal subpattern assignment}
\acro{ICI}{inter-carrier interference}
\acro{InF}{indoor factory}
\acro{IoT}{internet of things}
\acro{IFFT}{inverse fast Fourier transform}
\acro{ISAC}{integrated sensing and communication}
\acro{ISI}{inter-symbol interference}
\acro{i.i.d.}{independent, identically distributed}
\acro{JSC}{joint sensing and communication}
\acro{k-NN}{k-nearest neighbors}
\acro{LOS}{line-of-sight}
\acro{LTE}{long term evolution}
\acro{MAP}{maximum a posteriori}
\acro{MB}{multi-Bernoulli}
\acro{MBM}{multi-Bernoulli mixture}
\acro{MCL}{maximum coupling loss}
\acro{MPL}{maximum path loss}
\acro{MIL}{maximum isotropic loss}
\acro{MC}{Monte Carlo}
\acro{MIMO}{multiple-input multiple-output}
\acro{mMIMO}{massive multiple-input multiple-output}
\acro{MUSIC}{MUltiple SIgnal Classification}
\acro{MDL}{minimum description length}
\acro{MHT}{multi-hypothesis tracker}
\acro{mmWave}{millimeter-wave}
\acro{NR}{new radio}
\acro{NLoS}{non-line-of-sight}
\acro{OFDM}{orthogonal frequency division multiplexing}
\acro{OSPA}{optimal sub-pattern assignment}
\acro{p.d.f.}{probability density function}
\acro{PF}{particle filter}
\acro{PHD}{probability hypothesis density}
\acro{PSD}{power spectral density}
\acro{PPP}{Poisson point process}
\acro{QPSK}{quadrature phase shift keying}
\acro{QuaDRiGa}{QUAsi Deterministic RadIo channel GenerAtor}
\acro{RCS}{radar cross-section}
\acro{ReLU}{rectified linear unit}
\acro{RF}{radio frequency}
\acro{Rx}{receiver}
\acro{RFS}{random finite set}
\acro{RMSE}{root mean squared error}
\acro{r.v.}{random variable}
\acro{RRH}{remote radio head}
\acro{SSIR}{signal-to-self interference ratio}
\acro{SI}{self interference}
\acro{SNR}{signal-to-noise ratio}
\acro{SU}{single-user}
\acro{SCM}{sample covariance matrix}
\acro{SPF}{soft particle filter}
\acro{Tx}{transmitter}
\acro{TD}{time division}
\acro{TDD}{time-division duplexing}
\acro{TDOA}{time difference of arrival}
\acro{UMi}{urban micro}
\acro{UE}{user equipment}
\acro{ULA}{uniform linear array}
\acro{V2V}{vehicle-to-vehicle}
\end{acronym}
\DeclarePairedDelimiter{\nint}\lfloor\rceil
\DeclareMathOperator*{\argmax}{argmax}

\def\undertilde#1{\mathord{\vtop{\ialign{##\crcr
$\hfil\displaystyle{#1}\hfil$\crcr\noalign{\kern1.5pt\nointerlineskip}
$\hfil\tilde{}\hfil$\crcr\noalign{\kern1.5pt}}}}}

\begin{document}

\title{Multi-Base Station Cooperative Sensing\\ with AI-Aided Tracking}

\author{Elia~Favarelli,
        Elisabetta~Matricardi,
        Lorenzo~Pucci,
        Enrico~Paolini,
        Wen~Xu,
        Andrea~Giorgetti
\thanks{This work was supported by the CNIT National Laboratory WiLab and the WiLab-Huawei Joint Innovation Center.\\
\indent Elia~Favarelli, Elisabetta~Matricardi, Lorenzo~Pucci, Enrico~Paolini, and Andrea~Giorgetti are with the Department of Electrical, Electronic, and Information Engineering ``Guglielmo Marconi'' (DEI), University of Bologna, and the Wireless Communications Laboratory (WiLab), CNIT, Italy (e-mail: \{elia.favarelli, elisabetta.matricardi3, lorenzo.pucci3, e.paolini, andrea.giorgetti\}@unibo.it).\\
\indent Wen Xu is with Munich Research Center, Huawei Technologies Duesseldorf GmbH, Munich, Germany, (email: wen.dr.xu@huawei.com).} 
}

\markboth{}%
{}

\maketitle

\begin{abstract}
In this work, we investigate the performance of a \ac{JSC} network consisting of multiple \acp{BS} that cooperate through a \ac{FC} to exchange information about the sensed environment while concurrently establishing communication links with a set of \acp{UE}.
Each \ac{BS} within the network operates as a monostatic radar system, enabling comprehensive scanning of the monitored area and generating range-angle maps that provide information regarding the position of a group of heterogeneous objects. The acquired maps are subsequently fused in the \ac{FC}.
Then, a \ac{CNN} is employed to infer the category of the targets, e.g., pedestrians or vehicles, and such information is exploited by an adaptive clustering algorithm to group the detections originating from the same target more effectively.
Finally, two multi-target tracking algorithms, the \ac{PHD} filter and \ac{MBM} filter, are applied to estimate the state of the targets.
Numerical results demonstrated that our framework could provide remarkable sensing performance, achieving an \ac{OSPA} less than $\mathbf{60}\,$cm, while keeping communication services to \acp{UE} with a reduction of the communication capacity in the order of 10\% to 20\%.
The impact of the number of \acp{BS} engaged in sensing is also examined, and we show that in the specific case study, $\mathbf{3}$ \acp{BS} ensure a localization error below $\mathbf{1}\,\text{m}$.
\end{abstract}
\acresetall


\IEEEpeerreviewmaketitle

\section{Introduction}

\IEEEPARstart{T}{he} forthcoming generation of mobile radio networks is poised to offer a range of emerging functionalities, including innovative services. Notably, the ability to perform effective sensing using \ac{RF} signals has become feasible due to the evolution toward larger antenna arrays, namely \ac{mMIMO}, and higher frequency bands \cite{Tho:C21,Sch:C20}. The \ac{JSC} approach leverages existing communication infrastructure to provide sensing capabilities, offering advantages such as reduced costs and improved spectral and energy efficiency when compared to dedicated spectrum- and transceiver-dependent systems like radar \cite{ZhaWanLiu:23}. This convergence of sensing and communication systems envisioned for future networks will enable ubiquitous sensing services that rely on capturing reflections from non-collaborative objects, thus playing a critical role, e.g., in intelligent vehicular networks \cite{Wang:22}. Furthermore, the growing interest in sensing stems from its potential to support various applications, such as traffic monitoring, autonomous driving, safety in industrial environments, and environmental mapping \cite{Zhang:21,Cui:21IntegratingSA}.

The advent of \ac{mMIMO} technology in \ac{mmWave} bands facilitates the detection, tracking, and precise localization of pedestrians, vehicles, drones, and other moving objects in real-time scenarios \cite{liu:23integrated}. This enables the acquisition of range profiles of targets, a kind of target fingerprint, as scatterers in complex objects may be resolved into different range cells.
At the same time, the enormous advancement of \ac{AI}, and particularly image identification, has generated a vast and solid portfolio of solutions that could also be exploited in the field of \ac{ISAC} \cite{ChaGhaJos:18,KecAou:21,Tia:20}.

\begin{figure}[t]
    \centering    \includegraphics[width=1\columnwidth]{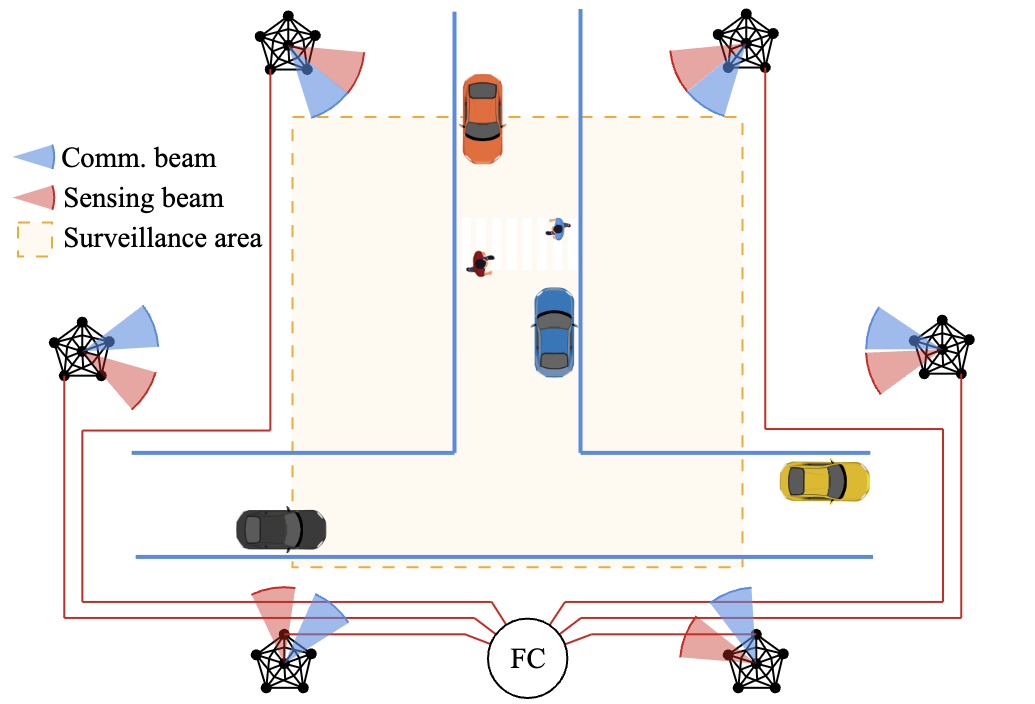}
    \caption{An urban scenario with $6$ monostatic \ac{JSC} \acp{BS} aiming at monitoring pedestrians (point-like targets) and vehicles (extended targets) in a surveillance area. \acp{BS} communicate with their \acp{UE} while simultaneously sensing the surrounding environment via dedicated sensing beams. The \ac{FC} collects measurements from the \acp{BS} via the backhaul network, fuses them to create likelihood maps, and performs detection, target identification, and multiple target tracking.
    }
    \label{fig:scenario}
\end{figure}

This work aims to investigate the possibility of using multi-sensor fusion techniques combined with multi-target tracking algorithms, to exploit range-angle radar maps obtained through a set of cooperating \acp{BS} with monostatic sensing capability and \ac{OFDM} signals.
The main contributions can be summarized as follows:
\begin{itemize}
    \item We propose a soft map fusion strategy based on range-angle maps obtained at each \ac{BS}.
    \item We present an \ac{AI}-based approach to infer the target category that is then exploited by an adaptive clustering methodology capable of managing point-like and extended targets.
    \item The adaptive clustering is then combined with tracking algorithms to perform target state estimation and prediction. Two different tracking algorithms, the \ac{PHD} and \ac{MBM} filter, are compared.
    \item We propose the \ac{OSPA} metric and aggregate downlink capacity to evaluate the sensing and communication capabilities.
    \item Finally, we investigate the impact of the number of cooperative \acp{BS} performing sensing on the localization and communication performance.
\end{itemize}

In this work, capital and lowercase boldface letters represent matrices and vectors, respectively; $\mathbf{D}_{q,t}$ stands for a matrix dependent on indexes $q$ and $t$, while $\mathbf{v}_{t,p}$ represents the $p$th column selected by the matrix $\mathbf{V}_t$. $\mathbf{I}_n$ is the $n\times n$ identity matrix; $\| \cdot \|_p$ stands for the $p$-norm;
$|\cdot |$ represents the cardinality of a set;
$\delta(\cdot)$ is the Dirac delta function; $\nint{\cdot}$ represents the round operator; $(\cdot)^c$ stands for conjugate; $\mathbf{x} \thicksim \mathcal{CN}(\mathbf{0},\boldsymbol{\Sigma})$ denotes a zero-mean circularly symmetric complex Gaussian random vector with covariance $\boldsymbol{\Sigma}$; and $\mathbf{x} \thicksim \mathcal{N}(\boldsymbol{\mu},\boldsymbol{\Sigma})$ denotes the real-valued Gaussian random vector with mean $\boldsymbol{\mu}$ and covariance $\boldsymbol{\Sigma}$.

The rest of the paper is organized as follows. Section~\ref{sec:system} presents the \ac{JSC} model. Section~\ref{sec:data} describes the data fusion strategy, target identification methodology, clustering scheme, and tracking algorithms. System performance is evaluated in Section~\ref{sec:results}, and conclusions are drawn in Section~\ref{sec:conclusion}.

\section{System Model}\label{sec:system}
This work considers a \ac{JSC} network, and a scenario, like the one portrayed in Figure~\ref{fig:scenario}.
In particular, the considered system consists of several monostatic \ac{JSC} \acp{BS} transmitting \ac{OFDM} signals at mmWave using \ac{mMIMO} technology. Each of these \acp{BS} is connected to an \ac{FC} via backhaul; the \ac{FC} allows them to cooperate in performing the detection and tracking of targets in the surveillance area. As shown later, the sensing task is accomplished through range-Doppler maps that each \ac{BS} can generate by scanning the environment using a dedicated sensing beam. Moreover, to ensure communication functionality, each \ac{BS} scans the environment for sensing and communicates with \acp{UE} in its respective cell using the same time-frequency resources via multiple beams.
To keep interference among the sensing beams of different \acp{BS} at a negligible level, we consider the proper use of \ac{FD} or \ac{TD} through coordination.

Each monostatic \ac{BS} is equipped with two separate \acp{ULA}, one for transmission and one for reception, with $N_\mathrm{T}$ and $N_\mathrm{R}$ antennas respectively, and both with a half-wavelength separation between the elements. In particular, the transmitted waveform is used for communication and sensing, while the sensing receiver (Rx) only collects backscattered signals.
More specifically, considering the downlink communication toward \acp{UE}, each \ac{BS} transmits frames consisting of $M$ \ac{OFDM} symbols and $K$ subcarriers, and the same signals are simultaneously used to sense the environment. A multi-beam radiation pattern is used to split the power between sensing and communication by exploiting spatial diversity, as explained later. In particular, each \ac{BS} uses a communication beam for the \ac{UE} while steering a sensing beam scanning the environment within the angular interval $[-\Theta_0,\Theta_0]$ with steps $\Delta\Theta$. In each sensing direction, a subset $M_\mathrm{s}<M$ of \ac{OFDM} symbols is collected by the Rx.

The \ac{OFDM} time-frequency grid containing the transmitted (complex) symbol for each sensing direction can be represented by a matrix $\mathbf{X}_\mathrm{s} \in \mathbb{C}^{K \times M_\mathrm{s}}$ with elements $x_{k}^{(m)}$, where $k$ is the subcarrier index and $m$ is the OFDM symbol (or time) index.

Starting from this grid, a precoding operation is performed on its elements with the beamformer $\mathbf{w}_\mathrm{T} \in \mathbb{C}^{N_\mathrm{T} \times 1}$ to map each complex symbol to each antenna and obtain the vector of the transmitted symbols $\tilde{\mathbf{x}}_{k}^{(m)} = \mathbf{w}_\mathrm{T} x_k^{(m)}$.
As previously mentioned, a multi-beam radiation pattern is considered at the Tx to split the total available power between the communication and sensing directions. Hence, the beamforming vector $\mathbf{w}_\mathrm{T}$ is defined as follows
\begin{equation}
\mathbf{w}_\mathrm{T} = \frac{\sqrt{P_\mathrm{T} G_\mathrm{T}^\mathrm{a}}}{N_\mathrm{T} }\left(\sqrt{\rho_\mathrm{p}}\mathbf{a}_\mathrm{T}^{c}(\theta_\mathrm{T,s}) + \sqrt{1-\rho_\mathrm{p}}\mathbf{a}_\mathrm{T}^{c}(\theta_\mathrm{T,c})\right)
\label{eq:BFvector}
\end{equation}
where $\rho_\mathrm{p} \in [0,1]$ is the fraction of power reserved for the sensing beam, $P_\mathrm{T}$ is the transmit power, $G_\mathrm{T}^\mathrm{a}$ is the transmit array gain along the beam steering direction, and $\mathbf{a}_\mathrm{T}(\theta_\mathrm{T,c}) \in \mathbb{C}^{N_\mathrm{T} \times 1}$ and $\mathbf{a}_\mathrm{T}(\theta_\mathrm{T,s}) \in \mathbb{C}^{N_\mathrm{T} \times 1}$ are the steering vectors associated with the communication and sensing directions, respectively, being $\theta_\mathrm{T,c}$ and $\theta_\mathrm{T,s}$ the respective \acp{DoD}.

Starting from the vector of the transmitted symbols $\tilde{\mathbf{x}}_{k}^{(m)}$, the vector $\tilde{\mathbf{y}}_{k}^{(m)} \in \mathbb{C}^{N_\mathrm{R} \times 1}$ of symbols received at each antenna, after \ac{OFDM} demodulation, is given by
\begin{equation}
    \tilde{\mathbf{y}}_k^{(m)} = \mathbf{H}_k^{(m)} \tilde{\mathbf{x}}_k^{(m)} + \tilde{\mathbf{n}}_k
    \label{eq:rx_signal}
\end{equation}
where $\mathbf{H}_k^{(m)} \in \mathbb{C}^{N_\mathrm{R} \times N_\mathrm{T}}$ is the channel matrix for the $m$th symbol and the $k$th subcarrier, which will be defined later, and $\tilde{\mathbf{n}}_k \sim \mathcal{CN}(\mathbf{0},\sigma_\mathrm{N}^2 \mathbf{I}_{N_\mathrm{R}})$ is the noise vector.\footnote{Both \ac{ICI} and \ac{ISI} are considered negligible.}

Spatial combining is then performed in the considered sensing direction, $\theta_\mathrm{R,s}=\theta_\mathrm{T,s}$, by using the receiving beamforming vector $\mathbf{w}_\mathrm{R} = \mathbf{a}_\mathrm{R}^c(\theta_\mathrm{R,s})$. This yields the grid of the received symbols $\mathbf{Y}_\mathrm{s} \in \mathbb{C}^{K \times M_\mathrm{s}}$, whose $(k,m)$ elements are defined as $y_k^{(m)} = \mathbf{w}_\mathrm{R}^T \tilde{\mathbf{y}}_k^{(m)}$. The received symbols grid collected in each sensing direction is then used to generate range-angle maps, as explained in Section~\ref{sec:range-angle-map}.

\subsection{Target Models}\label{sec:target-model}
This work considers both point-like targets, such as pedestrians, and extended targets, such as vehicles. Specifically, vehicles are represented by a model comprising $12$ reflection points. These include $4$ points to capture planar reflections originating from the front, back, and sides of the vehicle (characterized by a narrow visibility function and a substantial \ac{RCS}), $4$ points to account for the wheelhouses, and $4$ points to simulate the corners \cite{BuhrenYang:06,FavMatPuc:23,FavMatPuc2:23}.

Now, considering $L$ as the total number of reflections from both extended and point-like targets, the channel matrix already introduced in Equation~\eqref{eq:rx_signal} is given by
\begin{equation}\label{eq:channel-matrix}
    \mathbf{H}_k^{(m)} = \sum_{l = 1}^{L} \beta_l e^{j2\pi m T_\mathrm{s} f_{\mathrm{D},l}}e^{-j2\pi k \Delta f \tau_l} \mathbf{a}_\mathrm{R}(\theta_l)\mathbf{a}^T_\mathrm{T}(\theta_l)
\end{equation}
where $\Delta f = 1/T$ is the subcarrier spacing, $T_\mathrm{s} = T + T_\mathrm{cp}$ is the total \ac{OFDM} symbol duration including the cyclic prefix time $T_\mathrm{cp}$. Additionally, $f_{\mathrm{D},l}$ refers to the Doppler shift, $\tau_l$ represents the round-trip delay, $\theta_l$ denotes the \ac{DoA}, and $\mathbf{a}_\mathrm{R}(\theta_l)$ represents the array response vector at the Rx for the $l$th backscattered signal. The complex term $\beta_l = \left|\beta_l\right|e^{j\phi_l}$ includes phase shift and attenuation along the $l$th propagation path.
The \ac{SNR} at each receiving antenna related to the $l$th reflection point (hence the sensing \ac{SNR}) becomes
\begin{equation}
\begin{split}
    \text{SNR}^{(\mathrm{s})}_l & = \rho_\mathrm{p} \cdot \gamma_l \cdot  \frac{P_\mathrm{T} G_\mathrm{T}^\mathrm{a} G_\mathrm{R}}{\sigma_\mathrm{N}^2} \left|\beta_l\right|^2 \\
    & = \rho_\mathrm{p} \cdot \gamma_l \cdot  \frac{P_\mathrm{T} G_\mathrm{T}^\mathrm{a} G_\mathrm{R}}{N_0 K \Delta f} \frac{c^2 \sigma_{\mathrm{rcs},l}}{(4\pi)^3 f_\mathrm{c}^2 d_l^4}
\end{split}
    \label{eq:SNR-path-loss}
\end{equation}
where $G_\mathrm{R}$ represents the gain of a single antenna element at the Rx, $\gamma_l=|\mathrm{AF}(\theta_\mathrm{T,s}-\theta_l)|^2 \in [0,1]$ denotes the normalized array gain at the Tx, which considers the imperfect alignment between the sensing direction and the target \ac{DoA}, $N_0$ is the one-sided noise \ac{PSD} at Rx, $d_l$ represents the distance between the $l$th reflection point and the \ac{BS}, $\sigma_{\mathrm{rcs},l}$ corresponds to the \ac{RCS}, $f_\mathrm{c}$ is the carrier frequency and $c$ is the speed of light. 

The \acp{RCS} $\sigma_{\mathrm{rcs},l}$ of scatterers for both pedestrians and vehicles are random and modeled according to a Swerling~I type distribution whose mean value, $\bar{\sigma}_{\mathrm{rcs}}$, can be found in Table~\ref{tab:rcs} \cite{Sko:B08}. It is important to note that the number of backscattered signals $L$ depends on the relative angular position with respect to the \ac{BS} and varies over time according to a visibility function \cite{BuhrenYang:06}, as objects are moving.

\begin{table}[t]
\caption{Average \ac{RCS} for different point reflections}
\begin{center}
\renewcommand{\arraystretch}{1.1}
\begin{tabular}{|c|c|c|}
    \hline
    Reflection & $\bar{\sigma}_{\mathrm{rcs}}\,[\mathrm{m}^2]$ \\ 
    \hline
    \hline
    Pedestrian & 1 \\ 
    \hline
    Surfaces & 20 \\ 
    Wheelhouses & 0 \\ 
    Corners & 5 \\ 
    \hline
\end{tabular}
\end{center}\label{tab:rcs}
\end{table}

\begin{figure*}[t]
    \centering
    \includegraphics[width=1.8\columnwidth]{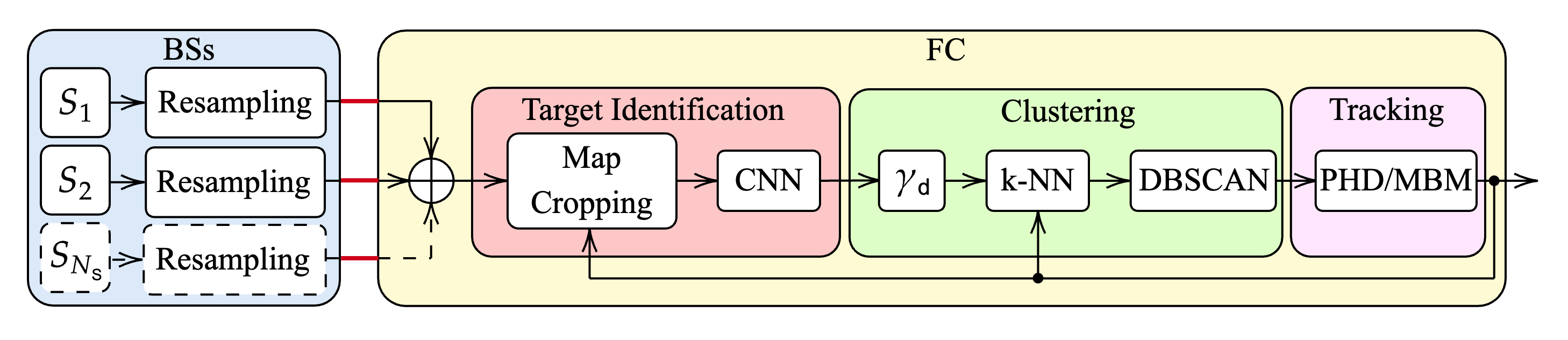}
    \caption{Block diagram of the sensing processing chain exploiting \ac{BS} cooperation, target classification, and target-specific tracking. The \acp{BS} scan the environment generating range-angle maps and resample them according to a predefined grid. Resampled range-angle maps are then shared with the \ac{FC} and fused in a single map. Target identification is performed at the \ac{FC} through map cropping and classification (red block). Then clustering is performed to merge detections generated by the same target (green block). Finally, tracking algorithms perform target state estimation (pink block).}
    \label{fig:block}
\end{figure*}

\subsection{Measurement Model}\label{sec:range-angle-map}
As mentioned, each \ac{BS} detects objects in the environment by scanning using a multi-beam pattern defined in Equation~\eqref{eq:BFvector}. Specifically, the communication beam is directed toward a \ac{UE}, while the sensing direction changes over time, sequentially pointing toward various directions following a predefined angular increment. In each direction, a set of $M_\mathrm{s}$ \ac{OFDM} symbols is collected to form the grid of received symbols $\mathbf{Y}_\mathrm{s}$, which is then used to obtain a range-angle map. The period required to complete a full scan, denoted as $T_\mathrm{scan}$, depends on the chosen number of sensing directions and on the symbol duration $T_\mathrm{s}$. Once all the symbols are acquired and assembled into the matrix $\mathbf{Y}_\mathrm{s}$, the first step involves an element-wise division between $\mathbf{Y}_\mathrm{s}$ and $\mathbf{X}_\mathrm{s}$, an operation often indicated as reciprocal filtering \cite{PucPaoGio:J22,Rod:J23}. This division aims to eliminate the influence of the transmitted symbols and generate a new matrix denoted as $\mathbf{G}_\mathrm{s}$. Subsequently, a double-periodogram is performed on the rows and columns of $\mathbf{G}_\mathrm{s}$ to obtain a range-Doppler map \cite{PucPaoGio:J22}. From this map, a range-angle map $\mathbf{D}_{q,t}$ is derived at the $q$th \ac{BS} and $t$th scan by selecting the column of the periodogram with the maximum value and uniquely associating it with the corresponding scan direction.\footnote{The estimation of target parameters turns out to be a frequency estimation problem; hence, since the periodogram represents (asymptotically) the log-likelihood, the column with the maximum value is selected.}

\section{Data Fusion, Target Classification, and Target-Oriented Processing}\label{sec:data}
According to the block diagram depicted in Figure~\ref{fig:block}, each \ac{BS} exchanges the range-angle map denoted as $\mathbf{D}_{q,t}$ with the \ac{FC}. The \ac{FC} employs a linear uniform grid, with resolution $\Delta_\mathrm{x}$ and $\Delta_\mathrm{y}$ (with $N_\mathrm{x}$ and $N_\mathrm{y}$ points) as a baseline. The received maps are rotated and translated according to the specific \ac{BS} position and \ac{ULA} orientation, and resampled at the baseline grid to ensure consistent map fusion. Subsequently, the resampled range-angle maps, represented as $\overline{\mathbf{D}}_{q,t}$, are combined via element-wise summation to yield the soft map $\mathbf{L}_{t} = \sum_{q=1}^{N_\mathrm{s}}\overline{\mathbf{D}}_{q,t}$, where $N_\mathrm{s}$ is the number of \acp{BS} performing sensing.\footnote{Since $\mathbf{L}_{t}$ are obtained via periodogram estimation, they can be interpreted as target log-likelihood maps, hence their summation results from noise independence among \ac{BS}.}

\subsection{Target Identification}
Each target exhibits a different reflection pattern related to its geometrical shape and \ac{RCS}, namely its reflection fingerprint. To this end, a \ac{CNN} is adopted to infer the target category (pedestrian or vehicle) directly from the resampled and fused soft maps $\mathbf{L}_{t}$ which contain such information.

Following Figure~\ref{fig:block}, a first step named image cropping is required to isolate each target from the others. A square window with side $W_\mathrm{size}$ pixels is selected to frame each target. Such windows are centered in the predicted target position at time $t$, inferred by the tracking algorithms exploiting information extracted during the previous time step $t-1$.
To generate the training set for the \ac{CNN}, we consider a scenario where actual target positions and categories are known.
To increase the classifier performance and robustness in the presence of imperfect target state predictions, which result in a misalignment between targets and relative frames, during training, the real target position is perturbed, adding Gaussian noise (which acts as a random displacement) with standard deviation $\sigma_\mathrm{w}$ on both $x$ and $y$ directions.
This solution leads to more accurate target classification, reducing the generalization error.
At the end of the training phase, the \ac{CNN} can infer the target category in real time and in a different scenario.

\subsection{Adaptive Clustering}
A three-step clustering procedure is employed to extract detections from the soft maps, enabling effective handling of extended and point-like objects (refer to Figure~\ref{fig:block} green block for a visual representation of the clustering procedure).
The main steps of the proposed strategy can be summarized as follows:
\begin{enumerate}
    \item An excision filter is implemented with threshold $\gamma_\mathrm{d}$, to remove points with low values from the $\mathbf{L}_{t}$ maps which are likely produced by noise. 
    \item A \ac{k-NN} algorithm with $k=1$ and adaptive gate $\xi_\mathrm{k}$ (to ignore residual points distant from each target) are applied to cluster data that likely belong to a previously detected target \cite{WatBorKat:16}. It is important to highlight that the parameter $\xi_\mathrm{k}$ can be adapted and varied depending on the target category. Section~\ref{sec:results} compares the solution with fixed values of $\xi_\mathrm{k}$ and the adaptive solution.
    \item The remaining points (i.e., map points larger than $\gamma_\mathrm{d}$ and outside the gate $\xi_\mathrm{k}$) are clustered through the \ac{DBSCAN} algorithm, with a maximum distance between points belonging to the same cluster $\xi_\mathrm{d}$, and a minimum number of points to form a cluster $N_\mathrm{d}$ \cite{EstKriXia:96}.
\end{enumerate}

Finally, each cluster centroid is stored in the matrix $\mathbf{Z}_{t}$, representing target detections extracted from the soft maps.

\subsection{Tracking Algorithms}
For all the tracking algorithms, we adopt the following state vector to represent the state of each target
\begin{equation}
    \mathbf{s}_{t,n} = \big(s_{t,n,\mathrm{x}},  s_{t,n,\mathrm{y}},  s_{t,n,v_\mathrm{x}},  s_{t,n,v_\mathrm{y}}\big)^\mathrm{T}
\end{equation}
where $t$ and $n$ are the time and target indexes.
The first two elements of the vector correspond to the target position coordinates, while the last two represent the target velocity components.
To update the target position coordinates, we use the information extracted from the map, while the target velocity components are inferred by considering both the previous target position at time $t-1$ and the current target position.

The \ac{PHD} filter is a widely adopted algorithm in literature \cite{MahRon:07,VobMaw:06}. One possible implementation suggests approximating the target intensity function as a \ac{GM} with a predefined number of components, which takes the following form
\begin{equation}
    D_{t-1|t-1}(\mathbf{x}) = \sum_{h=1}^{\mathcal{H}_{t-1|t-1}} w_{t-1|t-1}^{(h)} \mathcal{N}_\mathbf{x} (\boldsymbol{\mu}_{t-1|t-1}^{(h)},\mathbf{P}_{t-1|t-1}^{(h)})
    \label{eq:PHD}
\end{equation}
where $\mathbf{x}$ is a generic \ac{RFS}, $\mathcal{H}_{t-1|t-1}$ represents the number of Gaussian components in the intensity function, $w_{t-1|t-1}^{(h)}$ is the $h$th component weight, and $\boldsymbol{\mu}_{t-1|t-1}^{(h)}$ and $\mathbf{P}_{t-1|t-1}^{(h)}$ represent mean and covariance of the considered component.
The intensity function can be interpreted as an atypical \ac{p.d.f.} whose integral returns the estimated number of targets in the scenario.

The prediction step infers the intensity function in the consecutive time step, i.e., $D_{t|t-1}(\mathbf{x})$, through a linear Kalman predictor \cite{LiLiRan:15}.
During prediction, the probability of survival $P_\mathrm{s}$ is considered constant, so are the transition matrix $\mathbf{F}$ and the process noise covariance matrix $\mathbf{Q}$; the last one represents the motion uncertainty.
A set of $\mathcal{B}$ birth components is added to the predicted intensity function $D_{t|t-1}(\mathbf{x})$ to represent the possibility of new targets spawning in the surveillance area.
The total number of components after prediction is then $\mathcal{H}_{t|t-1} = \mathcal{H}_{t-1|t-1} + \mathcal{B}$.

In the update step, the predicted components are updated through the Kalman update equations, as in \cite{LiLiRan:15}, with the measurements $\mathbf{Z}_t$ extracted from the maps $\mathbf{L}_t$. During this step, the detection probability $P_\mathrm{d}$ is considered constant, and the covariance matrix $\mathbf{R}_t$ for each measurement is estimated from the selected map detection points, as will be highlighted in Equation~\eqref{eq:R}.
The overall amount of components in the posterior can be written as $\mathcal{H}_{t|t} = \mathcal{H}_{t|t-1}(M_t + 1)$, where $M_t$ denotes the number of measurements at time instant $t$.

To estimate the number of targets from the \ac{PHD} posterior, it is enough to sum the weight of the components and round it to the closest integer
\begin{equation}
    \widehat{N}_{\mathrm{obj}} = \nint*{\sum_{h=1}^{\mathcal{H}_{t|t}} w_{t|t}^{(h)}}
\end{equation}
while for the $n$th target state estimation, we extract the mean value of the $n$th most likely component
\begin{equation}
    \hat{\mathbf{s}}_{t,n} = \argmax_{ w_{t|t}^{(h)}} \boldsymbol{\mu}_{t|t}^{(h)}
\end{equation}

The \ac{MBM} filter is an alternative to the \ac{PHD} filter for multiple target tracking problems that exploit the association probability between measurements and targets \cite{GarXiaGra:19,GarWilGra:18}.
The \ac{MBM} filter is used to approximate the target multi-object \ac{p.d.f.}
\begin{equation}
    \mathrm{MBM}_{t-1|t-1}(\mathbf{x}) = \sum_{g = 1}^{\mathcal{G}_{t-1|t-1}} w^{(g)}_{t-1|t-1} \mathrm{MB}^{(g)}_{t-1|t-1}(\mathbf{x})
    \label{eq:MBM}
\end{equation}
where $\mathcal{G}_{t-1|t-1}$ represents the number of \ac{MB} components or global hypothesis in the \ac{MBM} distribution, and $w^{(g)}_{t-1|t-1}$ stands for the $g$th \ac{MBM} component weight.
The \ac{MB} distribution in Equation~\eqref{eq:MBM} can be written as follows
\begin{equation}
   \mathrm{MB}^{(g)}_{t-1|t-1}(\mathbf{x}) = \sum_{\biguplus \mathbf{x}_l = \mathbf{x}} \prod_{l = 1}^{\mathcal{L}^{(g)}_{t-1|t-1}}  \mathrm{B}^{(g,l)}_{t-1|t-1}(\mathbf{x}_l)
   \label{eq:MB}
\end{equation}
where $\mathcal{L}^{(g)}_{t-1|t-1}$ represents the number of Bernoulli components or local hypothesis in the \ac{MB} distribution, and the summation is performed for all the possible unions of mutually disjoint \ac{RFS} that generate $\mathbf{x}$, which means to evaluate all the possible data associations between measurements and targets \cite{GarWilGra:18}.
The single Bernoulli component in Equation~\eqref{eq:MB} can be written as
\begin{equation}
    \mathrm{B}^{(g,l)}_{t-1|t-1}(\mathbf{x}_l) = r_{t-1|t-1}^{(g,l)} \mathcal{N}_{\mathbf{x}_l}(\boldsymbol{\mu}_{t-1|t-1}^{(g,l)},\mathbf{P}_{t-1|t-1}^{(g,l)})
\end{equation}
where $r_{t-1|t-1}^{(g,l)}$ represents the existence probability of the $l$th local hypothesis in the $g$th global hypothesis, $\boldsymbol{\mu}_{t-1|t-1}^{(h)}$ and $\mathbf{P}_{t-1|t-1}^{(h)}$ represent the mean and the covariance of the considered component, respectively.

During prediction, linear Kalman prediction is performed again to infer the parameters in the consecutive time step. To account for new spawning objects, a set of $\mathcal{B}$ Bernoulli components is added to each global hypothesis. For both algorithms, to exploit the prior information about the environment, the components are generated following the scenario layout, i.e., the number of hypotheses, their mean value, covariance, and weight are based on the lanes and crosswalk positions in the environment.
The overall number of components after the prediction step can be evaluated as $(\mathcal{L}_{t-1|t-1}+\mathcal{B})\mathcal{G}_{t-1|t-1}$.

In the update phase, a linear Kalman update is performed to derive the updated parameters, considering the most likely association between measurements and targets \cite{GarXiaGra:19}.
Estimations $\hat{\mathbf{s}}_{t,n}$ are then extracted from the posterior distribution, considering the mean value $\boldsymbol{\mu}_{t|t}^{(i,j)}$ of the \ac{MB} components with existence probability $r_{t|t}^{(i,j)} \geq \gamma_\mathrm{e}$ from the \ac{MBM} component with highest probability $w_{t|t}^{(i)}$.

\subsection{Motion and Measurement Model}
To model clutter measurements representing false alarm detection extracted by the clustering procedure, a \ac{PPP} is considered, whose intensity is defined as $\lambda_\mathrm{c}$.

Target death is modeled through a constant probability of survival $P_\mathrm{s}$. During prediction, if a component is associated with a missed detection, its weight is multiplied by a factor proportional to $P_\mathrm{s}$, which means that consecutive missed detections lead to unlikely target state components.

A linear prediction model is selected to track the behavior of both extended and point-like targets.
This is justified by the low value of $T_\mathrm{scan}$ compared to the target velocity, which allows to approximate target motions as piecewise linear among consecutive acquisitions. The corresponding transition matrix and process noise covariance matrix are
\begin{equation}
    \mathbf{F} = 
    \begin{bmatrix}
        1 & 0 & T_{\mathrm{scan}} & 0 \\
        0 & 1 & 0 & T_{\mathrm{scan}} \\
        0 & 0 & 1 & 0 \\
        0 & 0 & 0 & 1
    \end{bmatrix}
    \label{eq:F}
\end{equation}

\begin{equation}
    \mathbf{Q} = \alpha_\mathrm{q}\,T_{\mathrm{scan}}\cdot\mathbf{I}_4
\end{equation}
where $\alpha_\mathrm{q}$ is a parameter that represents the prediction uncertainty about the target motion.

In this work, only position information about the targets is estimated through measurements, while consecutive position measurements are used to infer the velocity.\footnote{Although possible we consider the \ac{BS} do not estimate target Doppler.} With these assumptions, the following measurement matrix is considered
\begin{equation}
    \mathbf{H} = 
    \begin{bmatrix}
        1 & 0 & 0 & 0 \\
        0 & 1 & 0 & 0
    \end{bmatrix}.
    \label{eq:H}
\end{equation}

Because of the high-resolution maps, multiple detections (closely spaced map pixels) from each target are generated, leading to a non-diagonal measurement covariance matrix. Thus such a matrix needs to be estimated. Let us define the set of map points $\mathbf{L}_t^{(\mathbf{z}_{t,m})}$ extracted after clustering, specified by $\mathbf{z}_{t,m}$, from the measurement matrix (map) $\mathbf{L}_t$. Indicating with $\mathbf{V}_t^{(\mathbf{z}_{t,m})}$, the $2\times N_{\mathbf{z}_{t,m}}$ matrix containing the pixel coordinates relative to $\mathbf{L}_t^{(\mathbf{z}_{t,m})}$, the sample covariance measurement matrix can be calculated as
\begin{equation}
    \mathbf{R}_t = \frac{1}{N_{\mathbf{z}_{t,m}}-1} \sum _{{p=1}}^{N_{\mathbf{z}_{t,m}}}(\mathbf{v}_{t,p}^{(\mathbf{z}_{t,m})} - \mathbf{z}_{t,m})(\mathbf{v}_{t,p}^{(\mathbf{z}_{t,m})} - \mathbf{z}_{t,m})^{{\mathrm {T}}}
    \label{eq:R}
\end{equation}
where $N_{\mathbf{z}_{t,m}}$ represents the number of map points associated to the $m$th measurement $\mathbf{z}_{t,m}$, and $\mathbf{v}_{t,p}^{(\mathbf{z}_{t,m})}$ stands for the $p$th map point coordinates in the matrix $\mathbf{V}_t^{(\mathbf{z}_{t,m})}$.

\subsection{Post-Processing}
A set of post-processing procedures are implemented to manage the complexity of the algorithms and ensure good estimation accuracy.
In the \ac{PHD} filter, pruning, capping, and merging are implemented sequentially to reduce the number of components in the posterior intensity function.
Pruning removes all the components in the posterior whose weights $w_{t|t}^{(h)}$ are under a predefined threshold $\gamma_\mathrm{p}$ \cite{VobMaw:06}. Then, capping is realized on the remaining components selecting the $\gamma_\mathrm{q}$ components with the greatest $w_{t|t}^{(h)}$, by fixing the maximum number of components in the posterior to $\gamma_\mathrm{q}$ \cite{GarXiaGra:19}.

Finally, on the remaining components in the set $\zeta_\mathrm{m}$, merging of those whose average distance, defined in the following equation, is lower than a predefined threshold $\gamma_{\mathrm{s}}$ is performed:
\begin{equation}
d(\boldsymbol{\mu}_{t|t}^{(i)},\boldsymbol{\mu}_{t|t}^{(j)}) =  \|\boldsymbol{\mu}_{t|t}^{(i)}-\boldsymbol{\mu}_{t|t}^{(j)}\|_2
\end{equation}
where weights, mean, and covariance are updated as follows:
\begin{align*}
    w_{t|t}^{(k)} &= \sum_{i\in\boldsymbol{\zeta}_\mathrm{m}}w_{t|t}^{(i)} \\
    \boldsymbol{\mu}_{t|t}^{(k)} &= \sum_{i\in\boldsymbol{\zeta}_\mathrm{m}} w_{t|t}^{(i)}\boldsymbol{\mu}_{t|t}^{(i)} \\
    \mathbf{P}_{t|t}^{(k)} &= \sum_{i\in\boldsymbol{\zeta}_\mathrm{m}} w_{t|t}^{(i)}\mathbf{P}_{t|t}^{(i)} + (\boldsymbol{\mu}_{t|t}^{(i)}-\boldsymbol{\mu}_{t|t}^{(k)})(\boldsymbol{\mu}_{t|t}^{(i)}-\boldsymbol{\mu}_{t|t}^{(k)})^\mathrm{T}
\end{align*}
where $k$ represents the new index assigned to the derived component, and $i$ represents the index of the merged component.

In the \ac{MBM} filter, during the update phase, a gate for eligible data association allows pruning all the weak association hypotheses with $w_{t|t}^{(k)} < \xi_\mathrm{a}$. Both the \ac{MBM} and the \ac{MB} components are pruned with the threshold $\gamma_\mathrm{g}$ and $\gamma_\mathrm{l}$, respectively. Then, the residual \ac{MBM} components are capped with a threshold $\gamma_\mathrm{c}$.
To increase the estimation accuracy, in the most likely \ac{MBM} components, the \ac{MB} components closer than $\gamma_\mathrm{m}$ are merged as previously described.
\section{Numerical Results}\label{sec:results}

\subsection{Performance Metrics}
From the communication perspective, the aggregate network capacity, intended as the sum rate of each \ac{BS} in the downlink, is considered to assess the communication performance. Considering $N_\mathrm{s}$ \acp{BS} dedicated for both sensing and communications among the $N_\mathrm{tot}$ available \acp{BS} (so $N_\mathrm{tot}-N_\mathrm{s}$ are for communication only) and a fraction of power dedicated for sensing $\rho_\mathrm{p}$ \eqref{eq:BFvector}, the overall aggregate network capacity can be written as
\begin{equation}
\begin{split}
C(\rho_{\mathrm{p}}) = (N_\mathrm{s} \, \Delta f \, K \, \log_2(1+(1-\rho_\mathrm{p}) \, \mathrm{SNR}^{(\mathrm{c})})\\
    +\, (N_\mathrm{tot}-N_\mathrm{s}) \, \Delta f \, K \, \log_2(1+\mathrm{SNR}^{(\mathrm{c})}))/N_\mathrm{tot}
\end{split}
\label{eq:Shannon}
\end{equation}
where $\mathrm{SNR}^{(\mathrm{c})}$ is the communication \ac{SNR} experienced by the users.\footnote{To keep the presentation of numerical results simple, we consider all the \acp{UE} experience the same \ac{SNR}.}

To evaluate the network localization capability, the \ac{OSPA} is selected as a single-value metric \cite{RahGarSve:17,BeaBaBa:17}
\begin{equation*}
    \begin{split}
        &\mathrm{OSPA} = \\
        &\sqrt[\leftroot{-1}\uproot{5}\mathrm{p}]{\frac{1}{N_\mathrm{c}}\biggl(\sum_{(i,j)\in\boldsymbol{\zeta}_\mathrm{g}^*}\hspace{-8pt}d(\mathbf{\underline{s}}_{t,i},\hat{\mathbf{\underline{s}}}_{t,j})^\mathrm{p} + \frac{\xi_\mathrm{g}^\mathrm{p}}{2}(|\mathbf{\underline{S}}_t|+ |\hat{\mathbf{\underline{S}}}_t|-2|\boldsymbol{\zeta}_\mathrm{g}^*|)\biggr)}
    \end{split}
\end{equation*}
where $\hat{\mathbf{\underline{S}}}_t$ contains the position coordinates for all the estimated targets in the scenario representing the first two rows of $\hat{\mathbf{{S}}}_t$ inferred by the algorithms, and $N_\mathrm{c}$ is the number of elements in the \ac{OSPA} metric given by $N_\mathrm{c} = |\mathbf{\underline{S}}_t|+ |\hat{\mathbf{\underline{S}}}_t|-|\boldsymbol{\zeta}_\mathrm{g}^*|$, where $|\,\cdot\,|$ represents the \ac{RFS} cardinality.
The parameter $\mathrm{p}$ is the \ac{OSPA} order, while $\xi_\mathrm{g}$ represents the \ac{OSPA} gate.
Estimations beyond the gate threshold $\xi_\mathrm{g}$ are categorized as false alarms, while actual target positions not linked to estimations within the gate are classified as missed detections.
%
The set $\boldsymbol{\zeta}_\mathrm{g}^*$ represents the best assignment between the estimated set of objects $\hat{\mathbf{\underline{S}}}_t$ and the ground truth one $\mathbf{\underline{S}}_t$. The best assignment is selected as the one that minimizes the \ac{OSPA} error.
Note that the first term in the \ac{OSPA} metric, i.e., $d(\mathbf{\underline{s}}_{t,i},\hat{\mathbf{\underline{s}}}_{t,j})^\mathrm{p}$, can be interpreted as the distance between the estimated target positions and the real ones (and if $p=2$, this term corresponds to the square error of position estimation).
The rest of the \ac{OSPA} metric can be rewritten as
\begin{equation}
    \frac{\xi_\mathrm{g}^\mathrm{p}}{2}(|\mathbf{\underline{S}}_t|- |\boldsymbol{\zeta}_\mathrm{g}^*|) + \frac{\xi_\mathrm{g}^\mathrm{p}}{2}(|\hat{\mathbf{\underline{S}}}_t|- |\boldsymbol{\zeta}_\mathrm{g}^*|)
\end{equation}
where the first term is proportional to the missed detections, while the second term is related to the false alarms.

Finally, to evaluate the target classification performance, we define the classification accuracy as
\begin{equation}
    \mathrm{Accuracy} = \frac{T_\mathrm{P} + T_\mathrm{N}}{T_\mathrm{P} + T_\mathrm{N} + F_\mathrm{P} + F_\mathrm{N}}
\end{equation}
where $T_\mathrm{P}$, $T_\mathrm{N}$, $F_\mathrm{P}$, and $F_\mathrm{N}$ stand for true positive, true negative, false positive, and false negative classifications, respectively.

\subsection{Parameter Setup}
We considered a scenario with $N_\mathrm{tot}=6$ \acp{BS}, and a set of $4$ extended and $4$ point-like targets. Pedestrians and vehicles move as a mix of constant turns, linear accelerated/decelerated, static, and uniform linear motions. The area monitored is of size $x \in [-20,\,20]\,$m and $y \in [-20,\,20]\,$m. The \acp{BS} are positioned on a 
circumference of radius $50\,$m centered in the surveillance area, with the axes normal to the \acp{ULA} pointing toward the center, a scanning aperture of $120$° (i.e., $\Theta_0=60$°), and scan step $\Delta \Theta=2.4$°.

The transmission parameters are: \ac{QPSK} modulation, $f_\mathrm{c} = 28\,\text{GHz}$, $\Delta f = 120\,\text{kHz}$, $K= 3168$ (i.e., about $400\,\text{MHz}$ bandwidth), $M = 1120$, and $M_\mathrm{s} = 112$. The \ac{EIRP} is set to $30\,$dBm, and the noise \ac{PSD} is $N_0 = 4\cdot 10^{-20}\,$W/Hz. All the $6$ \acp{BS} are equipped with $N_\mathrm{T} = N_\mathrm{R} = 50$ antennas.
At each \ac{BS}, the scan duration is $T_\mathrm{scan}=50\,$ms, and the overall scenario is monitored for $10\,$s, resulting in $N_\mathrm{m} = 200$ measurements (maps collected). The grid resolution for map fusion is set to $\Delta_\mathrm{x} = 0.1\,$m and $\Delta_\mathrm{y} = 0.1\,$m.
The fraction of power dedicated to sensing $\rho_\mathrm{p}$ is set to $0.3$; the same for all \acp{BS}.

For the target identification task, the window size is set to $W_\mathrm{size} = 6\,$m, and the position perturbation standard deviation is $\sigma_\mathrm{w} = 0.5\,$m. The \ac{CNN} is composed of a 2D convolutional layer with $20$ random mask filters of dimension $5 \times 5$ and \ac{ReLU} activation function, a consecutive 2D max pooling layer which performs a down-sampling of a factor $2$, and a fully connected layer with the softmax activation function whose output dimension is $2$, to map the extracted features in the target classes.

In the clustering algorithm, the detection threshold is set to $\gamma_\mathrm{d} = 2\cdot10^{-7}$. The \ac{k-NN} gate $\xi_\mathrm{k}$ is tested for values between $4$ and $6$. For \ac{DBSCAN}, the cutoff distance is $\xi_\mathrm{d} = 3$, and the minimum number of points to form clusters is set to $N_\mathrm{d} = 50$.

In both tracking algorithms, the clutter intensity is $\lambda_\mathrm{c} = 0.1$, the prediction uncertainty is $\alpha_\mathrm{q} = 5$, and the initial component covariance is set to $\mathbf{P} = 0.5\cdot\mathbf{I}_4$. The probabilities of detection and survival are $P_\mathrm{d} = 0.99$ and $P_\mathrm{s} = 0.9$, respectively. The merging threshold is set to $\gamma_\mathrm{m} = 5$.
In the \ac{PHD} filter, the pruning threshold of the components is $\gamma_\mathrm{p} = 100\cdot10^-6$, while the maximum number of components is fixed to $\gamma_\mathrm{q} = 10$.
In the \ac{MBM} filter, the pruning threshold on the probability of existence is $\gamma_\mathrm{l} = 100\cdot10^-6$ while the pruning threshold on the \ac{MBM} components is set to $\gamma_\mathrm{g} = 10^{-15}$. The maximum number of \ac{MBM} components is $\gamma_\mathrm{c} = 10$. The gate for the admissible associations is set to $\xi_\mathrm{a} = 14$. The existing threshold is $\gamma_\mathrm{e} = 0.99$.
For both the algorithms, the birth components for new appearing objects are initialized with covariance $\mathbf{P}^{(b)} = 0.1\cdot\mathbf{I}_4$, with position $\boldsymbol{\mu}^{(b)}$ reflecting the possible target spawn position. A recovery component is initialized centered in the scenario with covariance $\mathbf{P}^{(b)} = 5\cdot\mathbf{I}_4$.

\subsection{Target Classification Performance}
\begin{figure}[t]
    \centering    
    \includegraphics[width=1\columnwidth]{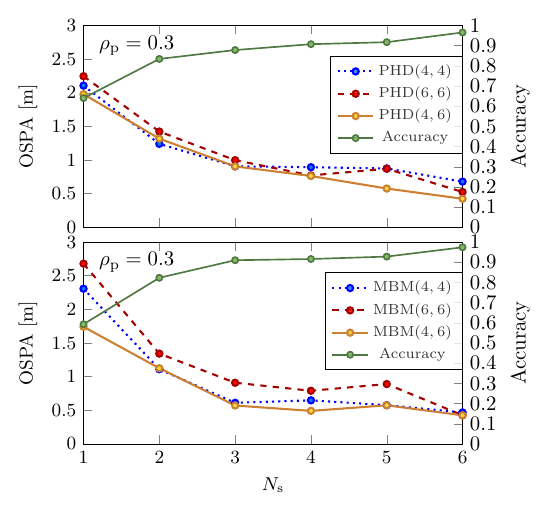}
    \caption{Localization performance and classification accuracy varying the number of \acp{BS} devoted for sensing $N_\mathrm{s}$ for the \ac{PHD} (top) and \ac{MBM} (bottom)}
    \label{fig:Ns}
\end{figure}

In Figure~\ref{fig:Ns}, the classification accuracy for $\rho_\mathrm{p} = 0.3$, and varying the number of \acp{BS} selected for sensing $N_\mathrm{s}$, is reported in green for both \ac{PHD} and \ac{MBM}, on the top and bottom plots, respectively.
It can be noticed that both the algorithms ensure a target classification accuracy greater than $0.85$ when the number of \acp{BS} is $N_\mathrm{s}\geq3$. Classification performance is highly influenced by the number of \acp{BS} adopted for sensing; reducing the number of \acp{BS} reduces the number of detected reflection points of extended targets (because of the reduction of spatial diversity), resulting in more similar target fingerprints between pedestrians and vehicles in the fused maps.
It is also interesting to notice that with $N_\mathrm{s} = 6$ \acp{BS}, the target classification accuracy is greater than $98\%$, representing a remarkable result.

\begin{figure}[t]
    \centering    
    \includegraphics[width=0.932\columnwidth]{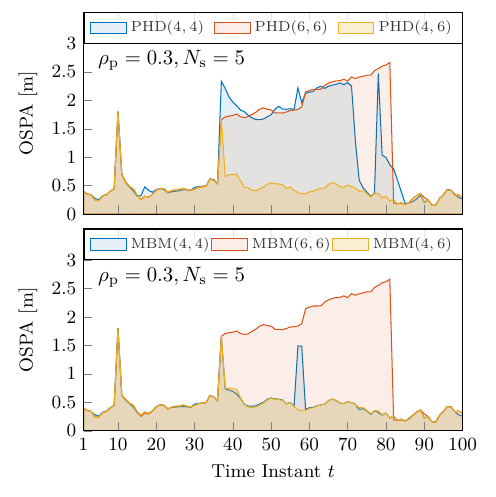}
    \caption{Localization performance over time for the \ac{PHD} (top) and \ac{MBM} (bottom)
    }
    \label{fig:time}
\end{figure}
\subsection{Sensing and Communication Performance}
In Figure~\ref{fig:Ns}, the \ac{OSPA} metric for $\rho_\mathrm{p} = 0.3$, and varying the number of sensors $N_\mathrm{s}$, is illustrated for \ac{PHD} and \ac{MBM}, on the top and bottom, respectively.
Blue dotted curves represent the algorithm performance with $\xi_\mathrm{k} = 4$ for both pedestrians and vehicles; red dashed curves refer to the performance with $\xi_\mathrm{k} = 6$ again for both pedestrians and vehicles. Solid yellow curves represent the performance of the AI-based solution, whose gates are adapted for pedestrians ($\xi_\mathrm{k} = 4$) and vehicles ($\xi_\mathrm{k} = 6$) based on the target identification.

As can be noticed, the adaptive gate achieves a lower localization error for both algorithms.
For the \ac{PHD} filter, the solution with adaptive gating presents an error lower than $1\,$m when the number of sensors is $N_\mathrm{s}\geq3$.
Similarly, the \ac{MBM} filter exhibits an \ac{OSPA} lower than $0.7\,$m considering $N_\mathrm{s}\geq3$.
The performance degradation experienced when $N_\mathrm{s}<3$ is due to target misclassification. In this case, the adaptive solution is affected by the mismatch between the real target classes and the estimated ones, resulting in an incorrect assignment of the gating parameter $\xi_\mathrm{k}$.

To emphasize the benefit produced by the adoption of adaptive gating (see Figure~\ref{fig:time}), the number of \acp{BS} devoted to sensing is fixed. At the same time, the \ac{OSPA} metric is reported over the first $100$ acquisitions.
Blue areas represent the \ac{OSPA} produced by a fixed gate $\xi_\mathrm{k} = 4$ for both pedestrians and vehicles, red areas refer to the solution with $\xi_\mathrm{k} = 6$, and yellow areas represent the adaptive solution.
It is important to highlight the increase in the localization performance thanks to adaptive gating for both algorithms, which results in reduced \ac{OSPA} peaks.

From a communication perspective, the \ac{BS} aggregate capacity is evaluated with Equation~\eqref{eq:Shannon}, considering $\rho_\mathrm{p} = 0.3$.
The worst case for communication is when all the \acp{BS} are performing joint communication and sensing, i.e., $N_\mathrm{s} = 6$. In this situation, the downlink capacity is $C = 0.9\,$Gbit/s.
On the contrary, without performing sensing ($N_\mathrm{s} = 0$), the downlink capacity can reach $C = 1.1\,$Gbit/s. As a compromise, using $3$ \acp{BS} for \ac{JSC} and $3$ for communication only, the downlink capacity can be maintained greater than $C = 1\,$Gbit/s.


\section{Conclusion}\label{sec:conclusion}
In this work, we presented a framework to perform \ac{JSC} with \ac{OFDM} waveforms exploiting cooperation and data fusion among \acp{BS} to improve localization performance. Furthermore, leveraging different target reflection fingerprints in the soft maps, we developed a \ac{CNN} classifier to identify the object type and adapt the multi-target tracking to the specific object type.

A three-step clustering strategy based on adaptive gating is proposed to manage point-like and extended targets and exploit target identification. Then, two multi-target tracking algorithms are used, the \ac{PHD} and \ac{MBM} filters, to track all the targets in the surveillance area.

The overall system is tested in a vehicular scenario with two types of targets, pedestrians and vehicles. To explore the communication/sensing trade-off, we investigated the sensing performance varying the number of cooperating \acp{BS}, considering that a fraction of transmit power is devoted to the sensing beams.

The system performance has been evaluated through the \ac{OSPA} metric, target classification accuracy, and communication performance via aggregate downlink capacity.
Numerical results show that adaptive gating aided by target identification performs better than the simpler target-agnostic solution when the target classification accuracy is greater than $90\%$.
For example, by choosing $N_\mathrm{s} = 3$ \acp{BS}, a classification accuracy around $0.9$ is reached, with an \ac{OSPA} error lower than $1\,$m for the \ac{PHD} filter and around $0.7\,$m for the \ac{MBM} filter, while also ensuring a downlink capacity greater than $1\,$Gbit/s.
With $N_\mathrm{s} = 6$ sensing \acp{BS}, a target classification accuracy larger than $98\%$ is reached, with a localization error lower than $0.7\,$m for both tracking algorithms, with a penalty on downlink capacity of 10\%, i.e., from $1\,$Gbit/s to $0.9\,$Gbit/s.






\bibliographystyle{IEEEtran}
\bibliography{IEEEabrv,bibliography}
\end{document}